\documentclass[10pt]{iopart}
\usepackage{graphicx}
\begin{document}
\title[Thermal fluctuations in superconducting phases with chiral $ d + id$ ]
{Thermal fluctuations in superconducting phases with chiral $ d + id $ and $ s $ symmetry 
on a triangular lattice }

\author{A G Groshev and A K Arzhnikov}

\address{Udmurt Federal Research Center of the Ural Branch of the Russian Academy of Sciences, T. Baramzinoy
st. 34, Izhevsk 426067, Russia}
\ead{groshev$_{-}$a.g@mail.ru}
\ead{arzhnikof@bk.ru}

\begin{abstract}
The behavior of thermal fluctuations of a superconducting order parameter with extended $ s $ 
and chiral $ d + id $ symmetry is investigated. The study is carried out on a triangular lattice 
within the framework of the quasi-two-dimensional single-band model with attraction between 
electrons at neighboring sites. The method of consistent  consideration of the order parameter 
fluctuations and the charge carrier scattering by fluctuations of coupled electron pairs, based 
on the theory of functional integration is used. The distribution functions of the phase fluctuation 
probabilities depending on temperature and charge carrier concentration are obtained. The temperature 
dependences of the amplitudes of the averaged superconducting order parameter are calculated. A phase 
diagram of superconducting states is constructed for the entire range of variation in the charge 
carrier concentration $ 0 <n <2 $. Near the boundaries of this range, topologically trivial 
superconducting states with extended $ s $ symmetry are realized, while a superconducting state with 
topologically nontrivial chiral $ d + id $ symmetry is realized between them. The calculated anomalous 
self-energies are compared with the experimental ones obtained using machine learning techniques.
\end{abstract}


\submitto{\JPCM}
\ioptwocol

\section{Introduction}

Recently, there has been an increased interest in studying superconductivity in layered materials with 
a triangular lattice, such as: sodium cobaltites $ Na_{x} CoO_{2}{\cdot y} H_{2} O $ intercalated with 
water, organic dielectrics $ k-(bis (ethylenedithio) $ $-tetrathiafulvalene )_{2} $ $ Cu_{2}(CN)_{3} $ 
and $ k-(BEDT-TTF)_{2} Cu_{2}(CN)_{3} $ \cite{Yamashita_2009}, $ In_{3} Cu_{2} VO_{9} $ \cite{Wu_2013}, 
$ SrPtAs $ \cite{Fischer_2014}, molybdenum disulfide $ MoS_{2} $ \cite{Yuan_2014}, bilayer graphene 
samples \cite{Vafek_2014}, and (111) bilayer perovskite transition metal oxides $ Na_{2}IrO_{3} $ 
($ Li_{2}IrO_{3}$) \cite{Okamoto_2013}. One of the reasons for this interest is a possibility to advance 
in solving the complicated long-standing problem of the nature of high-temperature superconductivity (HTSC). 
In addition, the triangular lattice is frustrated with respect to antiferromagnetic ordering, and 
superconductivity has a topologically nontrivial chiral $ d_{x^2-y^2} + id_{xy} $ symmetry, which is of 
interest per se. The presence of a topological phase opens the potential for a topological quantum transition 
under changes in the charge carrier concentration \cite{Zhou_2008,Valkov_2015} and for the appearance of 
Majorana states \cite{Lu_2013,Valkov1_2019,Valkov2_2019}. Furthemore, a number of materials are promising 
in practical application, see, for example, \cite{Ivanova_2009}. Since the compounds under study are highly 
anisotropic systems with effectively reduced (quasi-2D) dimension, it is important to take account of the 
increasing role of the order parameter (OP) fluctuations \cite{Emery_1995} when describing their properties. 
The allowance for these fluctuations substantially (by several times \cite{Groshev_2020}) reduces the 
temperature of superconducting transition $ T_c $ and, in some cases, leads to a change in the phase transition 
type \cite{Curty_2000,Bormannt_1994}. Usually, when considering superconducting properties, either these 
fluctuations are neglected, or only the amplitude or phase fluctuations of the OP 
$\Delta = \vert \Delta \vert \exp {(i \phi)} $ are taken into account. Sometimes, such approximations are 
justified \cite{Loktev}. However, in systems with a reduced dimension, it is important to take into account the 
amplitude and phase fluctuations of the OP \cite{Bormannt_1994,Groshev_2020} simultaneously. This is due to 
the fact that the phase and amplitude fluctuations turn out to be effectively related. Such a relationship was 
considered in \cite{Curty_2000,Curty_2003} in the framework of the variational approximation and in 
\cite{Groshev_2020} based on the self-consistent equations of the theory of continual integration in the 
coherent potential approximation. As indicated in \cite{Groshev_2020}, the approximations used in 
\cite{Curty_2000,Curty_2003} have some limitations. In particular, they do not take into account 
renormalization of the single-particle states resulting from the scattering of charge carriers on 
fluctuations of coupled electron pairs. Renormalization of the spectrum of single-particle states as a result 
of such scattering is described by the self-energy of the single-particle Green function, which defines the 
spectral density, so its consistent calculation seems necessary in explaining both the superconducting 
transition and the experimental data of angular resolution photoelectron spectroscopy (ARPES). Given the 
increased interest in superconducting states in layered materials with a triangular lattice, we believe it 
would be important to study the effect of a consistent consideration of the superconducting OP thermal 
fluctuations and the renormalization of single-particle states.

\section{Model and method of accounting for thermal fluctuations}

In this paper, we use the method proposed in \cite{Groshev_2020} which allows one to consistently take into 
account the renormalization of single-particle states and the effective relationship between the amplitude and 
phase fluctuations. This section presents only the main statements and features connected with the triangular 
lattice symmetry. We consider the one-band Hamiltonian of the $ t-V $ model with attraction between electrons 
at the nearest sites of a triangular lattice:
\begin {equation}
\label {eq:hamiltonian1}
{\hat {\cal H}} = \sum_{i,j,s} t_{ij} {\hat c}_{is}^{+}{\hat c}_{js}^{}-
\sum_{j}\mu {\hat n}_{j}-V \sum_ {j,\delta}{\hat n}_{j \uparrow} {\hat n}_{j+\delta \downarrow},
\end {equation}
where $t_ {ij} =-t $  are the matrix elements of electron jumps to the nearest sites; 
$\hat {c}_{js}^{+}(\hat {c}_{js}) $ are the operators of creation (annihilation) of an electron at site $ j $ 
with spin projection $ s $; $ n_{js}=\hat {c}_{js}^{+}\cdot\hat{c}_{js} $ is the operator of the number of 
electrons at site $ j $ with spin projection $ s $; $ n_ {j} $ is the operator of the total number of electrons 
at site $ j $; $ \mu $ is the chemical potential $; V $ is the parameter of inter-site attraction between electrons. 
The choice of such a Hamiltonian is justified by the presumed mechanisms of the appearance of superconductivity in 
HTSC. In this work, we do not specify the nature of the attraction between electrons at neighboring sites, 
assuming it to be due to either antiferromagnetic spin fluctuations \cite{Ivanova_2009}, or the state of 
resonating valence bonds \cite{Anderson_1987}, or other mechanisms (for example, the polaron one) which, in 
the simplest approximation, provide such an effective attraction 
\cite{Scalapino_1986,Schrieffer_1989,Izyumov_1999, Scalapino_2012}. Moreover, for simplicity, it is assumed 
that in the considered temperature range this attraction is weakly temperature-dependent, see \cite{Scalapino_1986}. 
The quasi-two-dimensional character of the compounds under study is taken into account by an approximation in which 
there are no fluctuations that destroy superconductivity in systems with dimension $ D\le 2 $. According to the 
Mermin-Wagner-Hohenberg theorem \cite{Mermin,Su_1997,Su_2007} in a strictly two-dimensional degenerate system, 
the long-range order is absent at any nonzero temperature and superconducting states can manifest themselves only 
in phase transitions of the Berezinskii-Kosterlitz-Thouless type \cite{Berezinskii}. The use of commutation 
transformations reduces the Hamiltonian (\ref {eq:hamiltonian1}) to the Hamiltonian of interacting electron pairs 
defined by the operators of creation 
$ {\hat O}_{j, \delta}^{+}={\hat c}_{j \uparrow}^{+} {\hat c}_{j+\delta\downarrow}^{+} $ and annihilation 
$ {\hat O}_{j, \delta}^{}={\hat c}_{j+\delta\downarrow}{\hat c} _ {j\uparrow} $ of an electron pair at site $ j $ 
and its nearest neighbor $ j + \delta $. This problem is solved within the method of continual integration, which 
proved effective earlier in studying the effect of temperature \cite{Groshev_2013} and atomic disorder 
\cite{Groshev_2018} on the magnetic phase separation and parameters of spiral magnetic structures in the framework 
of the quasi-two-dimensional one-band $ t-t '$ Hubbard model in the coherent potential approximation. In this paper, 
the problem under consideration is solved in the approximation of the average $t$-matrix, which reproduces quite 
well the results of the coherent potential approximation used in \cite{Groshev_2020}, but allows one to significantly 
reduce the computational cost. The Hubbard-Stratonovich transformation in the considered method allows us to reduce 
the problem of calculating the partition function of interacting electron pairs to that of calculating the partition 
function of independent electron pairs in the space of time-independent (in the static approximation) auxiliary 
fluctuating fields. The static approximation does not take into account the quantum fluctuations of the superconducting 
OP, which, in our opinion, are important only at sufficiently low temperatures. This is also evidenced by the estimate 
of the contribution of quantum fluctuations to the suppression of the superconducting transition temperature 
\cite{Emery_PRL}. Thus, for $T = 0$, the functional integration method reduces to the Hartree-Fock (HF) approximation. 
When calculating the partition function, the auxiliary fluctuating fields are considered in polar variables: the modulus 
$\Delta_{j,\delta}$ and the phase $\phi_{j,\delta}$, which determine the fluctuating complex OP 
$\Delta_{j,\delta}\cdot\exp(i\phi_{j,\delta})$ in site notation. In the ground state, thermal fluctuations 
of the superconducting OP are absent and $\phi_{j,\delta}=\alpha_{j,\delta} $, where $\alpha_{j,\delta} $ is 
the OP phase in the HF approximation, which determines its symmetry. Modern high-precision NMR data point to the 
spin-singlet Cooper pairing in the considered compounds \cite{Zheng_2006,Kobayashi_2006}. Therefore, in this work, 
we restrict ourselves to the study of singlet superconducting phases with extended $ s $ symmetry in which the 
superconducting gap depends on the 
wave vector according to the law $\Delta (k)\propto\cos(k_{1})+\cos(k_{2 })+\cos(k_{2}-k_{1}) $, and chiral 
$ d_{x^2-y^2} + id_{xy} $ - symmetry with dependence 
$ \Delta(k)\propto\cos(k_{1})+\exp(i2\pi/3)\cos(k_{2})+$ $\exp(-i2\pi/3)\cos(k_{2}-k_{1}) $\cite{Lee_2006}, where 
$ k_{1(2)} $ are the values of the wave vector along the basic reciprocal lattice vectors. Such symmetry types are 
admissible in the group-theoretical analysis of the singlet order parameter on a triangular lattice. When taking into 
account the interaction between electrons within the first coordination sphere, one should consider the OP in which only 
the phase depends on the nearest neighbors 
$\overline{\Delta}_{j,\delta}\cdot\exp(i\alpha_{j,\delta})=\overline{\Delta}\cdot\exp(i\alpha_{\delta}) $:
\begin{equation}
\label{eq:phase}
\alpha_{\delta}=
\left\{
\begin{array}{rcl}
0,\qquad\delta =\pm {\bf a_{1}},\\
\alpha,\qquad\delta =\pm {\bf a_{2}},\\
-\alpha,\qquad\delta =\pm ({\bf a_{2}}-{\bf a_{1}}),\\
\end{array}
\right.
\end{equation}
where $ {\bf a_{1}} $ and ${\bf a_{2}} $ are the basic vectors of the triangular lattice. The phase value 
$ \alpha = 0 $ corresponds to the $ s $-symmetry, and $\alpha = 2\pi /3 $ gives chiral $ d + id $-symmetry. 
To simplify the problem and reduce computational effort when calculating the integral over the amplitude 
field $\Delta_{} $, the “saddle point” approximation is used, in which the fluctuating field $\Delta_{} $ is 
replaced by its value at the “saddle point”  $\Delta_{}(\phi_{})$. Thus, in this approximation, the most probable 
amplitude and phase fluctuations turn out to be related. This approximation is valid when amplitude fluctuations 
become so much faster than phase fluctuations that the amplitude field has time to adjust to the phase distribution 
in an equilibrium manner. As shown in \cite{Groshev_2020}, this approximation works well over a wide temperature 
range up to $ T_{c} $. The minimum condition $\partial\Omega /\partial\Delta_{}=0$ of the 
thermodynamic potential $\Omega $ serves as an equation for finding the “saddle point”,  being at the same time a 
self-consistent equation for determining the superconducting OP amplitude  $\Delta (\phi)$. At $ T = 0 $ it 
coincides with the self-consistency equation for the superconducting OP of the mean field theory (HF, BCS). However, 
at finite temperatures in the method of continual integration, the solution $\Delta = 0 $ is lacking  (see 
\cite{Groshev_2020}), so to determine the temperature of the superconducting transition $ T_{c} $, the condition 
for the averaged OP to be zero $ \langle\Delta (\phi)\cdot\exp(i\phi)\rangle = 0 $ is used. Therefore, the transition 
to the normal state in the considered approach occurs as a result of the loss of phase coherence of the fluctuating 
complex OP. In consequence of the approximations made, the partition function per one electron pair is represented 
as an integral only over the phase field $\phi $. Together with the equations for determining the chemical potential 
$\mu $ and the self-energy $\Sigma $ in the approximation of the average $ t $-matrix (see Section 3), this set of 
self-consistent equations is solved by the iteration method. The solution with the minimum value of the thermodynamic 
potential determines the superconducting properties of the model under consideration.

\section{Green functions in the approximation of the average $ t $-matrix}

When solving the problem under study, it is convenient to use the representation of Nambu site matrices
\begin{equation}
\label{eq:Nambu_matrices}
\begin{array}{c}
\displaystyle
{\hat c}_{j\delta}(\tau)=
\left[
\begin{array}{cccc}
{\hat c}_{j\uparrow}(\tau)\\ 
{\hat c}_{j+\delta\downarrow}^{+}(\tau)
\end{array}
\right],\,
\\
{\hat c}_{j\delta}^{+}(\tau)=
\left[
\begin{array}{cccc}
{\hat c}_{j\uparrow}^{+}(\tau)& 
{\hat c}_{j+\delta\downarrow}(\tau)
\end{array}
\right],
\end{array}
\end{equation}
in which the Matsubara Green function and the fluctuating potential determining the perturbation of the system by 
thermal fluctuations are defined by the expressions
\begin{equation}
\label{eq:Nambu}
\begin{array}{c}
\displaystyle
G_{j\delta}(\tau-\tau')=-
\left\langle T_{\tau}{\hat c}_{j\delta}(\tau){\hat c}_{j\delta}^{+}(\tau')
\right\rangle =
\\
\displaystyle
=
\left[
\begin{array}{cccc}
G_{j, j}^{\uparrow\uparrow}(\tau-\tau')& 
G_{j, j+\delta}^{\uparrow\downarrow}(\tau-\tau')\\
G_{j+\delta, j}^{\downarrow\uparrow}(\tau-\tau')&
G_{j+\delta, j+\delta}^{\downarrow\downarrow}(\tau-\tau')
\end{array}
\right],
\\
\displaystyle
\Delta{\hat {\cal U}}(\Delta, \phi,\tau)=
\sum_{j,\delta}
{\hat c}_{j\delta}^{+}(\tau)
\Delta{\cal U}_{j\delta}
{\hat c}_{j\delta}(\tau),
\\
\displaystyle
\Delta{\cal U}_{j\delta}=
\left[
\begin{array}{cccc}
0&\Delta{\hat {\cal U}}_{j,j+\delta}^{\uparrow\downarrow}\\
\Delta{\hat {\cal U}}_{j+\delta ,j}^{\downarrow\uparrow}&
0
\end{array}
\right],
\\
\displaystyle
\Delta{\hat {\cal U}}_{j,j+\delta}^{\uparrow\downarrow}=
V\exp{\left(i\alpha_{\delta}\right)}\left[\overline{\Delta}-
\Delta(\phi)\exp{\left(i\phi_{}\right)}\right],
\\
\displaystyle
\!\Delta{\hat {\cal U}}_{j+\delta ,j}^{\downarrow\uparrow}=
V\exp{\left(-i\alpha_{\delta}\right)}\left[\overline{\Delta}-
\Delta(\phi)\exp{\left(-i\phi_{}\right)}\right],
\end{array}
\end{equation}
where $\overline{\Delta}$ is the average OP amplitude, $G_{j, j}^{\uparrow\uparrow}(\tau-\tau')$ and  
$G_{j+\delta,j+\delta}^{\downarrow\downarrow}(\tau-\tau')$ are the normal, and 
$G_{j+\delta,j}^{\downarrow\uparrow}(\tau-\tau')$ and $G_{j,j+\delta}^{\uparrow\downarrow}(\tau-\tau')$ 
the anomalous Matsubara Green functions defined by standard relations 
\begin{equation}
\label{eq:n_greens}
\begin{array}{c}
\displaystyle
G_{j,j'}^{\uparrow\uparrow}(\tau-\tau')=-
\left\langle 
T_{\tau}{\hat c}_{j\uparrow}(\tau){\hat c}_{j'\uparrow}^{+}(\tau')
\right\rangle ,
\\
\displaystyle
G_{j,j'}^{\downarrow\downarrow}(\tau-\tau')=-
\left\langle 
T_{\tau}{\hat c}_{j\downarrow}^{+}(\tau){\hat c}_{j'\downarrow}(\tau')
\right\rangle ,
\\
\displaystyle
G_{j,j'}^{\downarrow\uparrow}(\tau-\tau')=-
\left\langle 
T_{\tau}{\hat c}_{j\downarrow}^{+}(\tau){\hat c}_{j'\uparrow}^{+}(\tau')
\right\rangle ,
\\
\displaystyle
G_{j,j'}^{\uparrow\downarrow}(\tau-\tau')=-
\left\langle 
T_{\tau}{\hat c}_{j\uparrow}^{}(\tau){\hat c}_{j'\downarrow}^{}(\tau')
\right\rangle .
\end{array}
\end{equation}
The fluctuating potential $\Delta{\hat{\cal U}} $ introduced in (\ref{eq:Nambu}) to  account for thermal OP 
fluctuations determines the scattering matrix ($ t $ -matrix) in the Dyson site equation
\begin{equation}
\label{eq:Dyson1}
\begin{array}{c}
\displaystyle
G_{j\delta}(i\omega_{n})=
\\
\displaystyle
=G^{AV}_{j\delta}(i\omega_{n})+G^{AV}_{j\delta}(i\omega_{n})
\Delta{\cal U}_{j\delta}G_{j\delta}(i\omega_{n})=
\\
\displaystyle
=G^{AV}_{j\delta}(i\omega_{n})+G^{AV}_{j\delta}(i\omega_{n})T_{j\delta}(i\omega_{n})
G^{AV}_{j\delta}(i\omega_{n}),
\end{array}
\end{equation}
where $G^{AV}_{j\delta}(i \omega_{n})$ is the Fourier transform of the Matsubara Green function with averaged OP; 
$ T_{j \delta}(i \omega_{n}) $ is the Fourier transform of the electron pair scattering matrix; 
$\omega_{n}=(2n + 1)\pi T$ are the Matsubara frequencies for Fermi particles. Thus, when fluctuations are taken 
into account in the considered model (\ref{eq:hamiltonian1}), the problem of off-diagonal disorder arises in 
disordered systems. In this paper, this problem is solved within the framework of the two-site approximation 
of the average $ t $ -matrix, which significantly reduces the computational cost in comparison with the 
self-consistent approximation of the coherent potential. In the average $ t $-matrix approximation, the effective 
medium that preserves the full symmetry of the system under consideration and consists of electron pairs with 
effective parameters is determined by the self-energy in the Dyson site equation for the effective average 
$ F_{j \delta} (i \omega_{n})=\langle G_{j \delta} (i \omega_{n})\rangle $  Matsubara Green function 
\begin{equation}
\label{eq:Dyson}
\begin{array}{c}
\displaystyle
F_{j\delta}(i\omega_{n})=
\\
\displaystyle
=G^{AV}_{j\delta}(i\omega_{n})+G^{AV}_{j\delta}(i\omega_{n})\Sigma_{j\delta}(i\omega_{n})
F_{j\delta}(i\omega_{n}),
\end{array}
\end{equation}
where $ F_{j \delta}(i \omega_{n})$ is the Fourier transform of the effective Matsubara Green function; 
$\Sigma_{j \delta}(i \omega_{n})$ is the Fourier transform of the self-energy. Then from the matrix equations 
(\ref{eq:Dyson1}) and (\ref{eq:Dyson}) the self-energy $\Sigma_{j \delta}(i \omega_{n})$ is expressed in terms 
of the average $ t $-matrix 
\begin{equation}
\label{eq:Sigma}
\begin{array}{c}
\displaystyle
\Sigma_{j\delta}(i\omega_{n})=
\\
\displaystyle
=\left[1+\langle T_{j\delta}(i\omega_{n})\rangle G^{AV}_{j\delta}(i\omega_{n})\right]^{-1}
\langle T_{j\delta}(i\omega_{n})\rangle.
\end{array}
\end{equation}
The transition to the quasi-momentum representation occurs as a result of Fourier transformation of the Nambu 
site matrices. In this representation, the Hamiltonian of the system considered with averaged OP $ {\hat{\cal H}}_{AV} 
$ has the following form 
\begin{equation}
\label{eq:Nambu_k}
\begin{array}{c}
\displaystyle
{\hat {\cal H}}_{AV}(\overline{\Delta},\alpha)=
\frac{1}{N}\sum_{k}
{\hat c}_{k}^{+}
{\cal H}_{AV}(k)
{\hat c}_{k},
\\
\displaystyle
{\cal H}_{AV}(k)=
\left[
\begin{array}{cccc}
{\cal H}_{AV}^{\uparrow\uparrow}(k)&
{\cal H}_{AV}^{\uparrow\downarrow}(k)\\
{\cal H}_{AV}^{\downarrow\uparrow}(k)&
{\cal H}_{AV}^{\downarrow\downarrow}(k)
\end{array}
\right],
\\
\displaystyle
{\cal H}_{AV}^{\uparrow\uparrow}(k)=\varepsilon_{k}-\mu, 
{\cal H}_{AV}^{\downarrow\downarrow}(k)=-\varepsilon_{k}+\mu
\\
\displaystyle
{\cal H}_{AV}^{\uparrow\downarrow}(k)=-2V\overline{\Delta}V_{k}(\alpha),
\\ 
\displaystyle
{\cal H}_{AV}^{\downarrow\uparrow}(k)=\left({\cal H}_{AV}^{\uparrow\downarrow}(k)\right)^{*},
\\
\displaystyle
\varepsilon_{k}=-2t\left[\cos{k_{1}}+\cos{k_{2}}
+\cos{\left(k_{2}-k_{1}\right)}\right],
\\
\displaystyle
V_{k}(\alpha)=\cos{k_{1}}+\exp(i\alpha)\cos{k_{2}}+
\\
\displaystyle
+\exp(-i\alpha)\cos{(k_{2}-k_{1})},
\end{array}
\end{equation}
where ${\hat c}_{k}^{+}, {\hat c}_{k} $ are the Nambu matrices in quasi-momentum representation; 
$\varepsilon_{k} $ is the dispersion law of the electron energy on the triangular lattice with jumps 
within the first coordination sphere; $ V_{k}(\alpha) $ is the dispersion law of superconducting 
OP with symmetry given by the phase value $\alpha $. Thus, the Hamiltonian of the effective medium is 
determined by the expressions
\begin{equation}
\label{eq:H_eff}
\begin{array}{c}
\displaystyle
{\hat {\cal H}}_{eff}(i\omega_{n})={\hat{\cal H}}_{AV}+{\hat \Sigma(i\omega_{n})},
\\
\displaystyle
{\hat \Sigma(i\omega_{n})}=
\frac{1}{N}\sum_{k}
{\hat c}_{k}^{+}
\Sigma_k(E)
{\hat c}_{k},
\\
\displaystyle
\Sigma_k(i\omega_{n})=
\\
\left[
\begin{array}{cccc}
\Sigma^{\uparrow}(E)&
\Sigma^{\uparrow\downarrow}_{k}(i\omega_{n})-{\cal H}_{AV}^{\uparrow\downarrow}(k)
\\
\displaystyle
\Sigma^{\downarrow\uparrow}_{k}(i\omega_{n})-{\cal H}_{AV}^{\downarrow\uparrow}(k)&
\Sigma^{\downarrow}(i\omega_{n})
\end{array}
\right],
\\
\displaystyle
\Sigma^{\uparrow\downarrow}_{k}(i\omega_{n})=
2\Sigma^{\uparrow\downarrow}(i\omega_{n})V_{k}(\alpha),
\\
\displaystyle
\Sigma^{\downarrow\uparrow}_{k}(i\omega_{n})=
2\Sigma^{\uparrow\downarrow}(i\omega_{n})V_{k}^{*}(\alpha),
\end{array}
\end{equation}
where $ \Sigma_k(i \omega_{n}) $ is the self-energy (\ref{eq:Sigma}) in the quasi-momentum representation. 
To take into account only contributions with the considered symmetry types to the OP, it is necessary that 
the coefficient $\Sigma^{\uparrow\downarrow} (i \omega_{n})$ in the anomalous self-energy (ASE) (\ref{eq:H_eff}), 
defining the effective OP, be a real energy function, whereas the normal self-energies (NSE) 
$\Sigma^{\uparrow}(i \omega_{n})$ and $\Sigma^{\downarrow} (i \omega_{n})$ are in general complex functions. 
The effective Matsubara Green function (\ref{eq:Dyson}) in the quasi-momentum representation is defined through 
the effective medium Hamiltonian $ {\hat{\cal H}}_{eff}(i \omega_{n})$ (\ref{eq:H_eff})
\begin{equation}
\label{eq:F_k}
\begin{array}{c}
\displaystyle
F_k(i\omega_{n})=
\frac{1}{i\omega_{n}-{\cal H}_{eff}(k)}=
\\
\displaystyle
=
\left[
\begin{array}{cccc}
F_{k}^{\uparrow}(i\omega_{n})&
F_{k}^{\uparrow\downarrow}(i\omega_{n})\\
F_{k}^{\downarrow\uparrow}(i\omega_{n})&
F_{k}^{\downarrow}(i\omega_{n})
\end{array}
\right],
\\
\displaystyle
F_{k}^{\uparrow (\downarrow)}(i\omega_{n})=\frac{i\omega_{n}\pm\varepsilon_{k}\mp\mu-
\Sigma^{\downarrow (\uparrow)}(i\omega_{n})}
{\left(i\omega_{n}-E_{k}^{+})(i\omega_{n}-E_{k}^{-}\right)},
\\
\displaystyle
F_{k}^{\uparrow\downarrow (\downarrow\uparrow)}(i\omega_{n})=
\frac{\Sigma^{\uparrow\downarrow (\downarrow\uparrow)}_{k}(i\omega_{n})}
{\left(i\omega_{n}-E_{k}^{+})(i\omega_{n}-E_{k}^{-}\right)},
\\
\displaystyle
E_{k}^{\pm}=\left[\Sigma^{\uparrow}(i\omega_{n})+\Sigma^{\downarrow}(i\omega_{n})\right]/2\pm
\\
\displaystyle
\pm\left[\left(\varepsilon_{k}-\mu+\left[\Sigma^{\uparrow}(i\omega_{n})+
\Sigma^{\downarrow}(i\omega_{n})\right]/2
\right)^{2}+
\right.
\\
\left.
\displaystyle
+\Sigma^{\uparrow}(i\omega_{n})\Sigma^{\downarrow}(i\omega_{n})+
\mid\Sigma_{k}^{\uparrow\downarrow}(i\omega_{n})\mid^{2}\right]^{1/2}.
\end{array}
\end{equation}
To determine the self-energy from the Dyson equations (\ref{eq:Dyson1}) and (\ref{eq:Dyson}), the matrix 
elements of the Matsubara Green functions are required in the representation of the Nambu site matrices. 
Their explicit expressions can be obtained using the symmetry properties of the considered system under 
Fourier transformation. First of all, this is the property of the inversion symmetry of the dispersion laws 
of the electron energy and superconducting OP (\ref{eq:Nambu_k}) $F_{- k}(i \omega_{n})=F_{k}(i \omega_{n})$. 
In addition, in the triangular lattice with $ s $- and $ d + id $-symmetry, the off-diagonal matrix elements 
of the effective Matsubara Green function (\ref{eq:F_k}) in the quasi-momentum representation have the symmetry 
property  $F_{k_{1},k_{2}}^{\uparrow\downarrow (\downarrow\uparrow)}(i\omega_{n})=$ 
$\exp{\left(\pm i\alpha_{}\right)}F_{k_{2},k_{1}}^{\downarrow\uparrow (\uparrow\downarrow)}(i\omega_{n})$, 
$F_{k_{1},k_{2}}^{\uparrow\downarrow (\downarrow\uparrow)}(i\omega_{n})=$ 
$\exp{\left(\pm i\alpha_{}\right)}F_{k_{1}-k_{2},k_{1}}^{\downarrow\uparrow (\uparrow\downarrow)}(i\omega_{n})$ and 
$F_{k_{1},k_{2}}^{\uparrow\downarrow (\downarrow\uparrow)}(i\omega_{n})=$ 
$\exp{\left(\mp i\alpha_{}\right)}F_{k_{2},k_{2}-k_{1}}^{\downarrow\uparrow (\uparrow\downarrow)}(i\omega_{n})$. 
As a result, the matrix elements in the site representation obey the relations 
$F_{j,j\pm\delta}^{\uparrow\downarrow}(i\omega_{n})=$ 
$\exp{\left(i\alpha_{\delta}\right)}F_{j,j\pm{\bf a_{1}}}^{\uparrow\downarrow}(i\omega_{n})$, 
$F_{j\pm\delta ,j}^{\downarrow\uparrow}(i\omega_{n})=$ 
$\exp{\left(-i\alpha_{\delta}\right)}F_{j\pm{\bf a_{1}},j}^{\downarrow\uparrow}(i\omega_{n})$. It is easy to 
verify that this is also true for the matrix elements of the Matsubara Green functions of the system considered 
with averaged OP and off-diagonal matrix elements: $\Delta{\cal U}_{j,j\pm\delta}^{\uparrow\downarrow}=$ 
$\exp{\left(i\alpha_{\delta}\right)}\Delta{\cal U}_{j,j\pm{\bf a_{1}}}^{\uparrow\downarrow}$, 
$\Delta{\cal U}_{j\pm\delta,j}^{\downarrow\uparrow}=$ 
$\exp{\left(-i\alpha_{\delta}\right)}\Delta{\cal U}_{j\pm{\bf a_{1}},j}^{\downarrow\uparrow}$ and 
$\Sigma_{j,j\pm\delta}^{\uparrow\downarrow}=$ 
$\exp{\left(i\alpha_{\delta}\right)}\Sigma_{j,j\pm{\bf a_{1}}}^{\uparrow\downarrow}$,
$\Sigma_{j\pm\delta,j}^{\downarrow\uparrow}=$ 
$\exp{\left(-i\alpha_{\delta}\right)}\Sigma_{j\pm{\bf a_{1}},j}^{\downarrow\uparrow}$. Therefore, the explicit 
expressions for the matrix elements of the Green function in the representation of Nambu site matrices 
$ G_{j, \delta} (i \omega_ {n}) $, determined from the Dyson site equation (\ref{eq:Dyson1}), also have this 
property $G_{j,j\pm\delta}^{\uparrow\downarrow}(i\omega_{n})=$ 
$\exp{\left(i\alpha_{\delta}\right)}G_{j,j\pm{\bf a_{1}}}^{\uparrow\downarrow}(i\omega_{n})$, 
$G_{j\pm\delta ,j}^{\downarrow\uparrow}(i\omega_{n})=$ 
$\exp{\left(-i\alpha_{\delta}\right)}G_{j\pm{\bf a_{1}},j}^{\downarrow\uparrow}(i\omega_{n})$. This property 
allows the use of one self-consistent equation to determine the superconducting OP amplitude  $\Delta (\phi )$, 
because  it is not any more dependent  on the nearest neighbors:
\begin{equation}
\label{eq:delta_phi1}
\begin{array}{c}
\displaystyle
\Delta^{2}(\phi)-\frac{\Delta(\phi)K(\phi)}{2}-\frac{1}{2\beta V}=0,
\\
\displaystyle
K(\phi)=
\frac{1}{\beta}\sum_{n}\exp{\left(i\phi_{}\right)}
G_{j+{\bf a_{1}}, j}^{\downarrow\uparrow}(i\omega_{n})+
\\
\displaystyle
+\frac{1}{\beta}\sum_{n}\exp{\left(-i\phi_{}\right)}
G_{j,j+{\bf a_{1}}}^{\uparrow\downarrow}(i\omega_{n}).
\end{array}
\end{equation}
In the general case, for example, with an anisotropic hopping integral or interaction, to determine the 
superconducting OP amplitudes, it is necessary to solve a system of self-consistent equations.


\section{Results}

The calculations have been performed for the intersite interaction parameter $ V = t $. This is due to the 
fact that the value $V \simeq t $ was used to analyze the topological properties of superconducting cobaltites 
\cite{Zhou_2007,Zhou_2008}. In addition, among other values, $V = t $ was chosen when calculating the phase 
diagrams of superconducting states for the considered model on a square lattice without taking into account the 
OP fluctuations \cite{Timirgazin_2019}. The results of calculating the dependence of the amplitude of the 
averaged OP on the charge carrier concentration at temperature T = 0.0004t are shown in Fig.~1. It is seen 
that the largest part of the phase diagram is occupied by a superconducting region with chiral $ d + id $ 
symmetry, on both sides of which superconducting states with extended $ s $ symmetry are realized. The presence 
of contiguous superconducting regions with different symmetry suggests a first-order phase transition and phase 
separation between them, which contradicts the properties of superconducting states. For the model on a square 
lattice, this contradiction can be avoided by including in consideration an intermediate phase with 
$ s + id $-symmetry which provides a minimum of free energy and second-order phase transitions 
\cite{Timirgazin_2019}. For the model on a triangular lattice, the superconducting states substantially differ 
from those in the model on a square lattice. The point is that the chiral $ d + id $ superconducting phase in the 
model on triangular lattice is topologically nontrivial (has a finite value of the topological parameter $ Q $, 
see \cite{Zhou_2008}). Then, according to the results of this work, at sufficiently low temperatures, when there 
is no normal-state region between the superconducting regions with $ s $- and $ d + id $-symmetry, a change in the 
charge carrier concentration should result in the transition from a topologically trivial state with extended 
s-symmetry to a superconducting state with a topologically nontrivial chiral $ d + id $-symmetry. Besides, it is 
known \cite{Valkov_2015,Zhou_2008} that under changes in the charge carrier concentration or other model parameters, 
the superconducting state with a chiral $ d + id $ symmetry type admits a quantum topological transition which 
occurs when the nodal points of the OP cross the Fermi contour of the normal phase. Therefore, the solution of the 
problem of the first-order phase transition between superconducting states turns out to be more complicated for the 
model on a triangular lattice than on a square one. Note that in the framework of this model, the formation of a 
normal state at $ T = 0 $ is impossible, since any arbitrarily small attractive interaction makes the system 
unstable with respect to the Cooper pairing \cite{Cooper_1956}. Therefore, the appearance in Fig.~1. of a wide 
region in which superconducting states are destroyed by thermal fluctuations even at such a low temperature 
$ T = 0.0004t $ points to a small phase stiffness of such states in this concentration range. The results of 
calculating the temperature dependences of the averaged OP amplitude $\langle\Delta\rangle $ in superconducting 
regions with $ d + id $- (n $\simeq $ 1.49) and $s$-symmetry (n $\simeq $ 0.03) and (n $\simeq $ 1.88) are shown in 
Fig.~2. and Fig.~3, respectively. A comparison of the figures shows that the superconducting states with 
$ s $-symmetry are much more sensitive to thermal fluctuations than those with $ d + id $ symmetry.
\begin{figure}[h]
\begin{center}
\includegraphics[width=\linewidth]{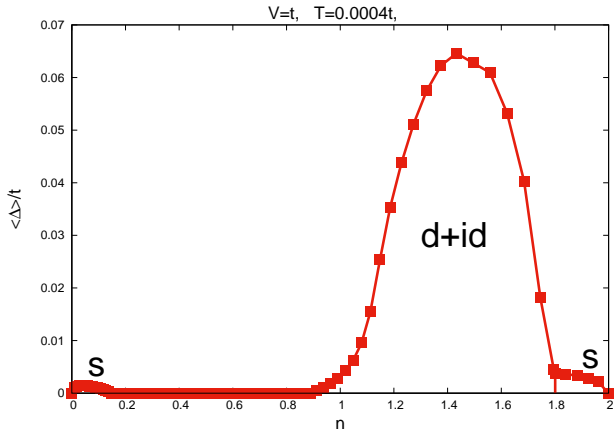}
\caption{ Dependence of the averaged OP amplitude on the charge carrier concentration at temperature T = 0.0004t.}
\end{center}
\end{figure}
\begin{figure}[h]
\begin{center}
\includegraphics[width=\linewidth]{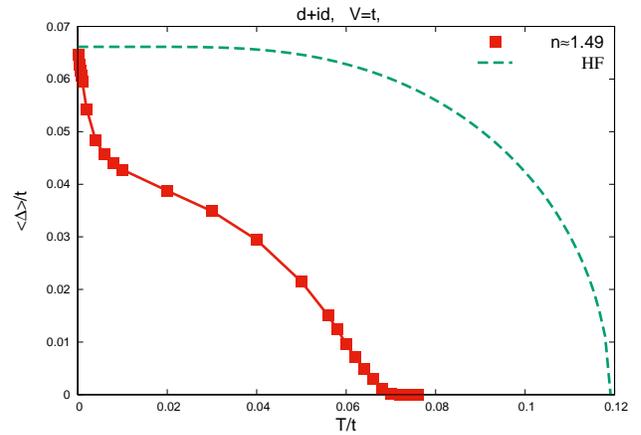}
\caption{ Temperature dependences of the amplitude of the averaged OP with $d + id$ symmetry
in different approximations.}
\end{center}
\end{figure}
\begin{figure}[h]
\begin{center}
\includegraphics[width=\linewidth]{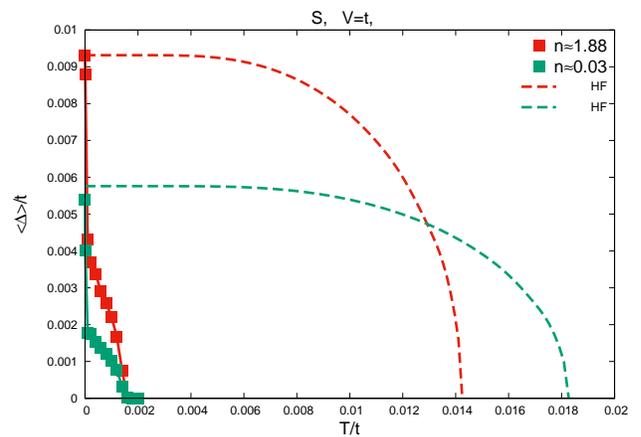}
\caption{ Temperature dependence of the amplitude of the averaged OP with extended s-symmetry.}
\end{center}
\end{figure}
Taking into account fluctuations for the states with $ s $-symmetry reduces the temperature of transition to 
the superconducting state with respect to the transition temperature calculated in the HF approximation by 11.2 
times for n $\simeq $ 0.03, and 8.7 times for n $\simeq $ 1.88, while for the states with $ d + id $-symmetry 
only by 1.7 times. Note also a significant variation with temperature of the behavior of the averaged OP amplitude. 
Another important parameter, which is often measured experimentally, is the ratio of the gap value at low 
temperatures to the superconducting transition temperature. Calculations of the ratio of the energy gap to the 
superconducting transition temperature in the HF approximation (without taking account of fluctuations) give the 
values $\Delta_{0}/T_{c}\simeq 1.1 $ for $ d + id $- , and $\Delta_{0}/T_{c}\simeq 0.6 $ (n $\simeq $ 0.03) and 
$\Delta_{0}/T_{c}\simeq 1.3 $ (n $\simeq $ 1.88) for extended s-symmetry. When fluctuations are taken into account, 
this ratio significantly increases $ \Delta_{0}/T_{c}\simeq 1.9 $ for $ d + id $- , and $\Delta_{0}/T_{c}\simeq 7.1$ 
(n $\simeq $ 0.03) and $\Delta_{0}/T_{c}\simeq 11.6 $ (n $\simeq $ 1.88) for extended s-symmetry (in the above 
notation $\Delta_{0}=2V\langle\Delta\rangle $). A larger value of the ratio $\Delta_{0}/T_{c} $ as compared to the 
BCS theory (by 5 times and more), is observed experimentally in layered chloronitride compounds with transition 
metals $ MNaCl $ $ (M = Zr, Hf) $ 
\begin{figure}[h]
\begin{center}
\includegraphics[width=\linewidth]{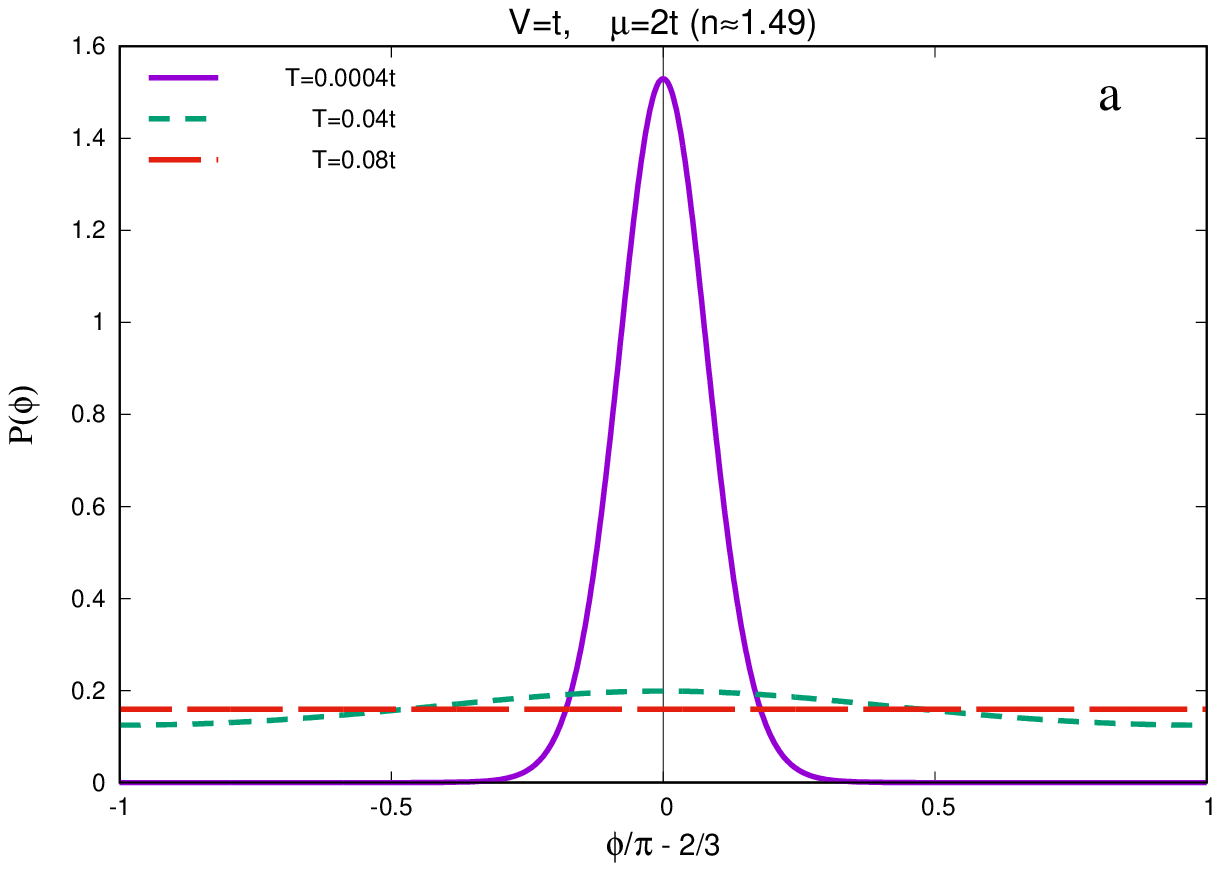}
\includegraphics[width=\linewidth]{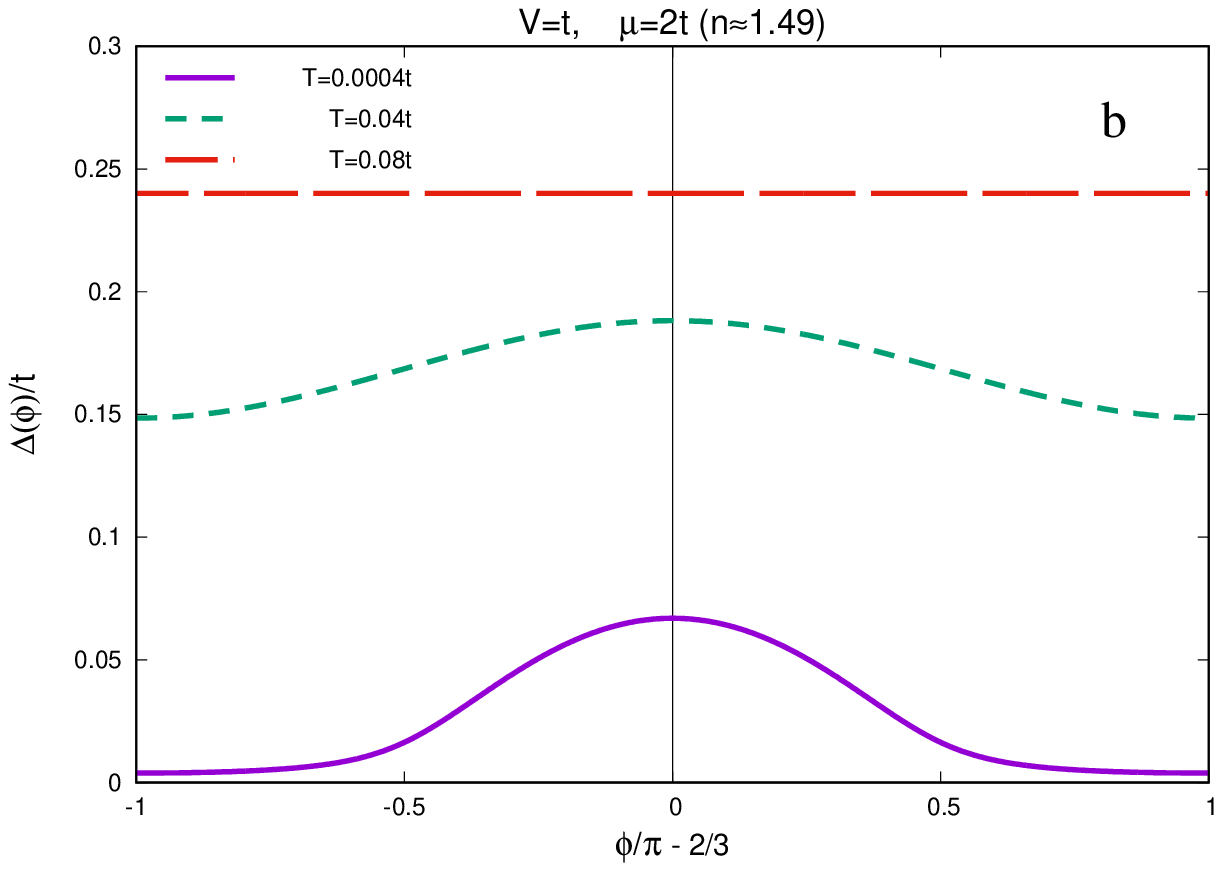}
\caption{(a) The probabilities of the distribution of phase fluctuations $ \phi $ of the 
superconducting OP with $d + id$ symmetry, and (b) the dependence of the amplitude of the OP with $d + id$ 
symmetry on $ \phi $.}
\end{center}
\end{figure}
\begin{figure}[h]
\begin{center}
\includegraphics[width=\linewidth]{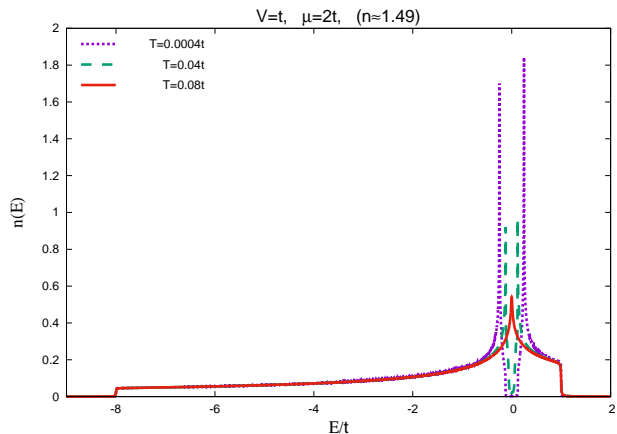}
\caption{Density of electron states calculated for the superconducting state with $d + id$ symmetry.}
\end{center}
\end{figure}
\begin{figure}[h]
\begin{center}
\includegraphics[width=\linewidth]{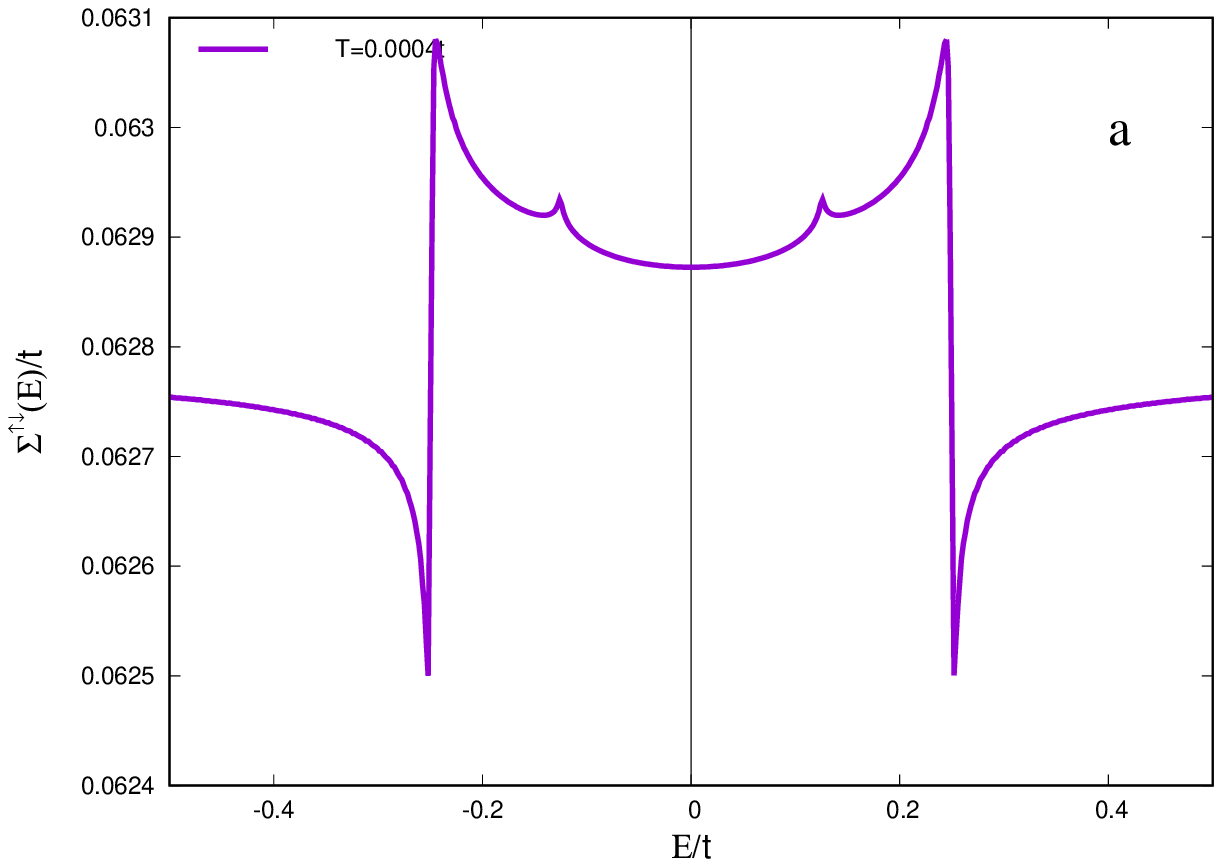}
\includegraphics[width=\linewidth]{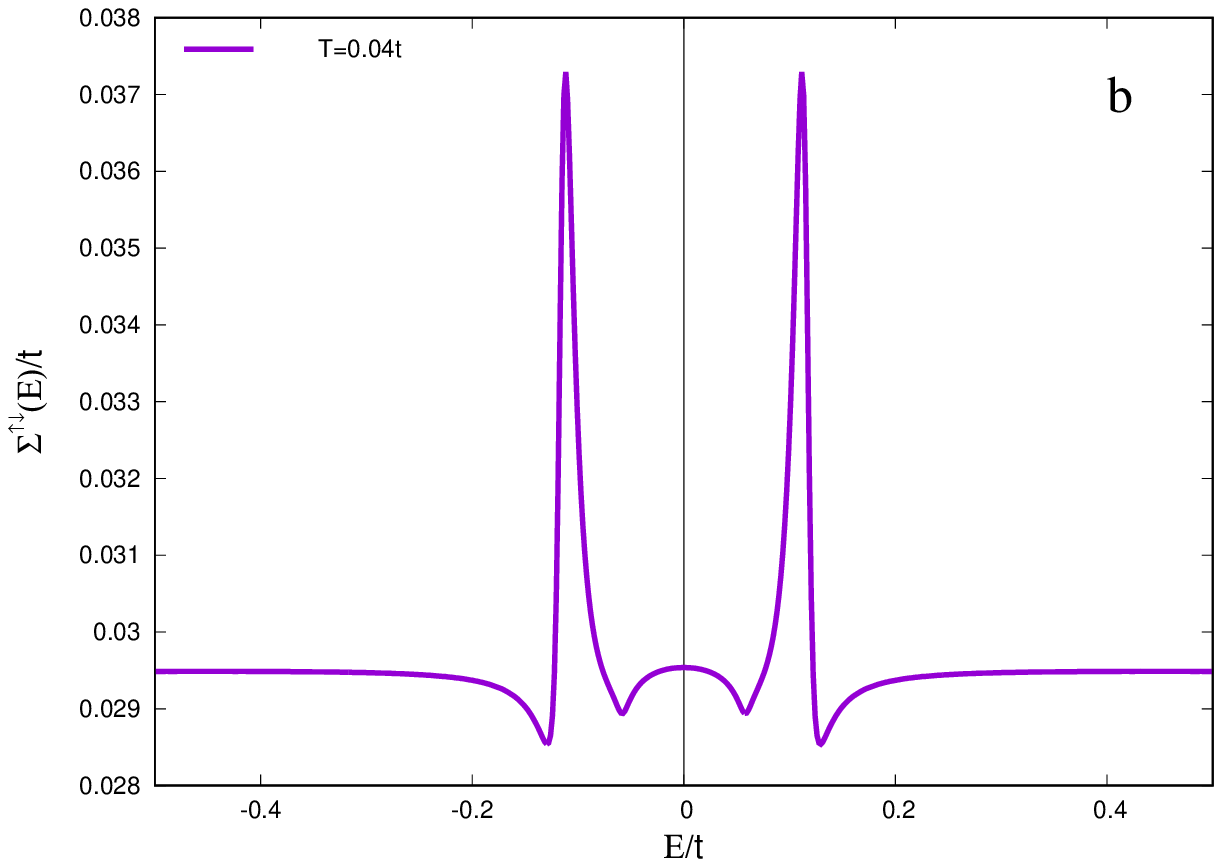}
\caption{Anomalous self-energy of the one-electron Green function, calculated for the superconducting state with 
$d + id$ symmetry for a) T=0.0004t and b) T=0.04t.}
\end{center}
\end{figure}
\cite{Ekino_2013} which are highly anisotropic effectively quasi-two-dimensional systems with a triangular lattice 
in each of the backbone planes. It should be emphasized that such a large ratio cannot be a consequence of the 
frustration of the triangular lattice, since the frustration effects diminish also the gap magnitude at low 
temperatures. The $\Delta_{0}/T_{c} $ ratio increases by about $\simeq $ 2.5 times for the $ d + id $-symmetry if 
the self-consistent method of coherent potential is used in the solution instead of the $ t $-matrix approximation 
\cite{Groshev_2020}.  In addition, we did not take into account the possible decrease in the effective attraction 
of electrons with increasing temperature \cite{Chen_2013}, which also leads to an increase of this ratio. Thus, the 
large experimental values of the ratio $\Delta_{0}/T_{c} $, in our opinion, are a consequence of thermal fluctuations 
of the superconducting OP, and are hardly determined by the details of the electron pairing mechanism. The results 
of a consistent calculation of the probabilities of the distribution of phase fluctuations and the most probable 
fluctuations of the amplitude, calculated in the “saddle point” approximation for three different temperatures, are 
shown in Fig.~4a, and Fig.~4b. It is seen that the most probable phase value $\phi =2\pi/3 $ corresponds to the 
$d + id $-symmetry. At the same time, considering the variation range of phase fluctuations $[-\pi /3, 5\pi /3] $ at 
finite temperatures, there is always a nonzero probability of finding the system in a superconducting state with the 
$ d-id $-symmetry (i.e., with the opposite chirality and $\phi =4\pi /3 $). Note an equally important result 
associated with a significant dependence of the most probable values of the OP amplitude on the phase  fluctuations. 
This means that approximations in which amplitude fluctuations are not taken into account, that is, the amplitude is 
a constant and does not depend on phase fluctuations, are not justified.

An important quantity affected by fluctuations of the superconducting OP in the self-consistent approach is the 
self-energy of the one-particle Green function, which determines the density of states and the energy dependence 
of the effective OP. The density of states for three different temperatures is shown in Fig.~5. It can be seen from 
the plots that the dip in the density of electron states decreases with increasing temperature and completely 
disappears at the point of the phase transition simultaneously with the loss of phase coherence of the OP. The energy 
dependence of the anomalous self-energy (ASE), which determines the energy dependence of the effective OP, is presented 
in Fig.~6. A comparison of the ASE dependences obtained in this work and in \cite{Groshev_2020} shows that the 
structure of the singularities in the approximations chosen is not determined by the lattice geometry, but correlates 
with the behavior of the density of electron states near the gap and its singularities associated with the formation 
of quasiparticle states. With increasing temperature, the distance between the ASE singularities decreases and vanishes 
together with the OP at the point of the phase transition. As the temperature decreases, the ASE becomes a constant, 
the value of which is equal to the OP in the HF approximation. The energy-dependent changes in the ASE amount to about 
10 $\% $ and may be observable experimentally. We are not aware of the methods of direct measurement of the ASE, but 
this contribution can be separated out from the ARPES data. In the paper \cite{Yamaji_2020}, using machine learning 
methods, it was possible to isolate the real part of the ASE depending on the energy for $Bi_2Sr_2CaCu_2O_8$ and 
$ Bi_2Sr_2CuO_6$. The obtained experimental dependence of the ASE \cite{Yamaji_2020} shows qualitative agreement with 
the theoretically calculated dependences presented in Fig.~6.

\section{Conclusion}

For a quasi-two-dimensional system with a triangular lattice, the effect of thermal fluctuations of the 
superconducting order parameter on the behavior of the superconducting gap, one-electron density of states, 
and phase transition temperature in singlet superconducting phases with extended s- and chiral $ d + id $-symmetry 
is studied. It is shown that the phase diagram of superconducting states, constructed in the whole range of 
variation of the charge carrier concentration $0<n<2 $, consists of two superconducting regions with a topologically 
trivial extended $ s $ - symmetry and a superconducting region with a topologically nontrivial chiral 
$ d + id $- symmetry between them. At sufficiently low temperatures, there is no normal-state region between the 
superconducting regions with $ s $- and $ d + id $- symmetry. In this case, as a result of variations in the charge 
carrier concentration, a phase transition should occur in the system from a topologically trivial superconducting 
state with an extended $ s $- symmetry to a superconducting state with a topologically nontrivial chiral 
$d + id $- symmetry. The elucidation of the type of this transition calls for further investigation. Taking into 
account the OP fluctuations significantly lowers the temperature of transition to the superconducting state, as a 
consequence, the ratio of the energy gap to the superconducting transition temperature increases several times. 
Then it turns out that superconducting states with extended $ s $- symmetry have a lower phase stiffness and, 
therefore, are more sensitive to thermal fluctuations than the states with chiral $ d + id $ symmetry. The 
calculated ASE values are semi-quantitatively consistent with the ASE values expressed from ARPES experimental data 
using machine learning techniques.

\ack
This study was supported by the financing program AAAA-A16-116021010082-8.

A G Groshev https://orcid.org/0000-0002-7389-5023

A K Arzhnikov https://orcid.org/0000-0002-8365-1962


\section*{References}
\begin{thebibliography}{37}

\bibitem{Yamashita_2009} Yamashita M, Nakata N, Kasahara Y,  Sasaki T,  Yoneyama N, 
 Kobayashi N,  Fujimoto S, Shibauchi T,  Matsuda Y 2009 Nature Phys. {\bf 5} 44 

\bibitem{Wu_2013}  Wu W, Scherer M M,  Honerkamp C, and  Le Hur K 2013 Phys. Rev. B 
{\bf 87} 094521

\bibitem{Fischer_2014}  Fischer M H,  Neupert T,  Platt C,  Schnyder A P,  Hanke W, 
 Goryo J, Thomale R and  Sigrist M 2014 Rev. B {\bf 89} 020509

\bibitem{Yuan_2014}  Yuan N F Q,  Mak K F, and  Law K T 2014 Phys. Rev. Lett. {\bf 113} 
097001

\bibitem{Vafek_2014}  Vafek O,  Murray J M, and  Cvetkovic V 2014 Phys. Rev. Lett. {\bf 112} 
147002

\bibitem{Okamoto_2013}  Okamoto S 2013 Phys. Rev. Lett. {\bf 110} 066403

\bibitem{Zhou_2008}  Zhou S,  Wang Z 2008 Phys. Rev. Lett. {\bf 100} 217002

\bibitem{Valkov_2015}  Val'kov V V,  Val'kova T A and  Mitskan V A 2015 JETP Letters {\bf 102} 361

\bibitem{Lu_2013}  Lu Y M  and  Wang Z 2013 Phys. Rev. Lett. {\bf 110} 096403

\bibitem{Valkov1_2019}  Val'kov V V  and  Zlotnikov A O 2019 JETP Letters {\bf 109} 736

\bibitem{Valkov2_2019}   Val'kov V V,  Mitskan V A,  Zlotnikov A O,  Shustin M S and 
 Aksenov S V JETP Letters {\bf 110} 140

\bibitem{Ivanova_2009}  Ivanova N B,  Ovchinnikov S G,  Korshunov M M,  Eremin I M and 
 Kazak N V 2009 Physics-Uspekhi {\bf 52} No. 8 789

\bibitem{Emery_1995}  Emery V J  and  Kivelson S A 1995 Nature {\bf 374} 434

\bibitem{Groshev_2020}  Groshev A G, and  Arzhnikov A K 2020 Journal of Experimental and Theoretical 
Physics {\bf 130} No. 2 247

\bibitem{Curty_2000}  Curty P and  Beck H 2000 Phys. Rev. Lett {\bf 85} 796 

\bibitem{Bormannt_1994}  Bormannt D and  Beck H 1994 J. Stat. Phys. {\bf 76} 361

\bibitem{Loktev}  Loktev V M,  Quick R M,  Sharapov S G 2001 Physics Reports {\bf 349} 1

\bibitem{Curty_2003}  Curty P and  Beck H 2003 Phys. Rev. Lett {\bf 91} 257002

\bibitem{Anderson_1987}  Anderson P W 1987 Science  {\bf 235} 1196

\bibitem{Scalapino_1986}  Scalapino D J,  Loh E, and  Hirsch J E 1986 Phys. Rev B {\bf 34}(11) 8190(R)

\bibitem{Schrieffer_1989}  Schrieffer J R,  Wen Х G, and  Zhang S С 1989 Phys. Rev B {\bf 39}(16) 11663

\bibitem{Izyumov_1999} Izyumov Y A 1999 Physics-Uspekhi {\bf 42}(3) 215

\bibitem{Scalapino_2012}  Scalapino D J 2012 Rev. Mod. Phys.{\bf 84} 1383

\bibitem{Mermin}  Mermin N D,  Wagner H 1966 Phys. Rev. Lett. {\bf 17} 1136; 
 Hohenberg P C 1967 Phys. Rev. {\bf 158} 383  Coleman S 1973 Commun. Math. Phys. 
{\bf 31} 264

\bibitem{Su_1997}  Su G,  Schadschneider A,  Zittartz J 1997  Phys. Lett. A {\bf 230} 99 

\bibitem{Su_2007}  Su G and  Suzuki M 1998 Phys. Rev B {\bf 58} 117

\bibitem{Berezinskii}  Berezinskii V L 1970 Zh. Eksp. Teor. Fiz. {\bf 59} 907; 
 Kosterlitz J,  Thouless D 1973 J. Phys. C {\bf 6} 1181

\bibitem{Groshev_2013}  Groshev A G and  Arzhnikov A K 2013 EPL {\bf 102} 57005

\bibitem{Groshev_2018}  Groshev A G and  Arzhnikov A K 2018 J. Phys.: Condens. Matter {\bf 30} 185801 (6pp)

\bibitem{Emery_PRL}  Emery V J,  Kivelson S A 1995 Phys. Rev. Lett {\bf 74} 3253

\bibitem{Zheng_2006}  Zheng G et al.  2006 Phys. Rev. B {\bf 73} 180503(R)

\bibitem{Kobayashi_2006}  Kobayashi Y et al.  2006 J. Phys. Soc. Jpn. {\bf 75} 074717

\bibitem{Zhou_2007}  Zhou Zc and  Wang Z 2007 Phys. Rev. Lett. {\bf 98} 226402

\bibitem{Timirgazin_2019}  Timirgazin M A,  Gilmutdinov V F,  Arzhnikov A K 2019 Physica C 
{\bf 557} 7 

\bibitem{Cooper_1956}  Cooper L 1956 Phys. Rev. {\bf 104} 1189

\bibitem{Ekino_2013}  Ekino T,  Sugimoto A,  Gabovich A M,  Zheng Z,  Yamanaka S 2013 Physica C {\bf 494} 89

\bibitem{Chen_2013}  Chen K S,  Meng Z Y, Yu U,  Yang S,  Jarrell M and  Moreno J 2013 Phys. Rev B {\bf 88} 
041103(R)

\bibitem{Lee_2006}  Lee T K,  Feng S1990 Phys. Rev. B {\bf 41} 11110

\bibitem{Yamaji_2020} Yamaji Y,  Yoshida T,  Fujimori A, and  Imada M 2020 arXiv:1903.08060.

\end{thebibliography}
\end{document}